\documentclass[aps,prb,twocolumn,amsmath,amssymb,superscriptaddress,scrartcl,eqsecnum,longbibliography]{revtex4-1}
\usepackage{extarrows}
\usepackage{slashed}
\usepackage{amsmath,amsfonts,amssymb,graphics,graphicx,epsfig,color,times,indentfirst,layout}
\usepackage{lipsum}
\usepackage{mathrsfs}
\usepackage[unicode=true,pdfusetitle,bookmarks=false,colorlinks=true,citecolor=black,urlcolor=black,linkcolor=black]{hyperref}
\usepackage{tikz}
\usepackage{tikz-cd}
\usetikzlibrary{arrows}
\usetikzlibrary{intersections}
\usetikzlibrary{shapes.geometric}

\def\be{\begin{equation}}
\def\ee{\end{equation}}

\begin{document}

\title{Entanglement evolution across a conformal interface
}

\date{\today}

\author{Xueda Wen}

\affiliation{Department of Physics, Massachusetts Institute of Technology, Cambridge, MA 02139, USA}

\author{Yuxuan Wang}
\affiliation{Institute for Condensed Matter Theory and Department of Physics,
University of Illinois at Urbana-Champaign, 1110 West Green St, Urbana IL 61801, USA}

\author{Shinsei Ryu}
\affiliation{James Franck Institute and Kadanoff Center for Theoretical Physics, University of Chicago, Illinois 60637, USA}

\begin{abstract}
For two dimensional conformal field theories in the ground state, it is known that a conformal interface along
the entanglement cut can suppress the entanglement entropy from $S_A\sim c\log L$ to $S_A\sim c_{\text{eff}}\log L$,
where $L$ is the length of the subsystem $A$, and $c_{\text{eff}}\in [0, c]$ is the effective central charge which
depends on the transmission property of the conformal interface.
In this work, by making use of conformal mappings, we show that a conformal interface has the same effect
on entanglement evolution in non-equilibrium cases, including global, local and certain inhomogeneous quantum quenches.
I.e., a conformal interface suppresses the time evolution of entanglement entropy by effectively
replacing the central charge
$c$ with $c_{\text{eff}}$, where $c_{\text{eff}}$ is exactly the same as that in the ground state case.
We confirm this conclusion by a numerical study on a critical fermion chain.
Furthermore, based on the quasi-particle picture, we conjecture that this conclusion holds for
an arbitrary quantum quench in CFTs, as long as the initial state can be described by a regularized
conformal boundary state.
\end{abstract}
\maketitle


\section{Introduction}

\begin{figure}[htp]
\includegraphics[width=3.00in]{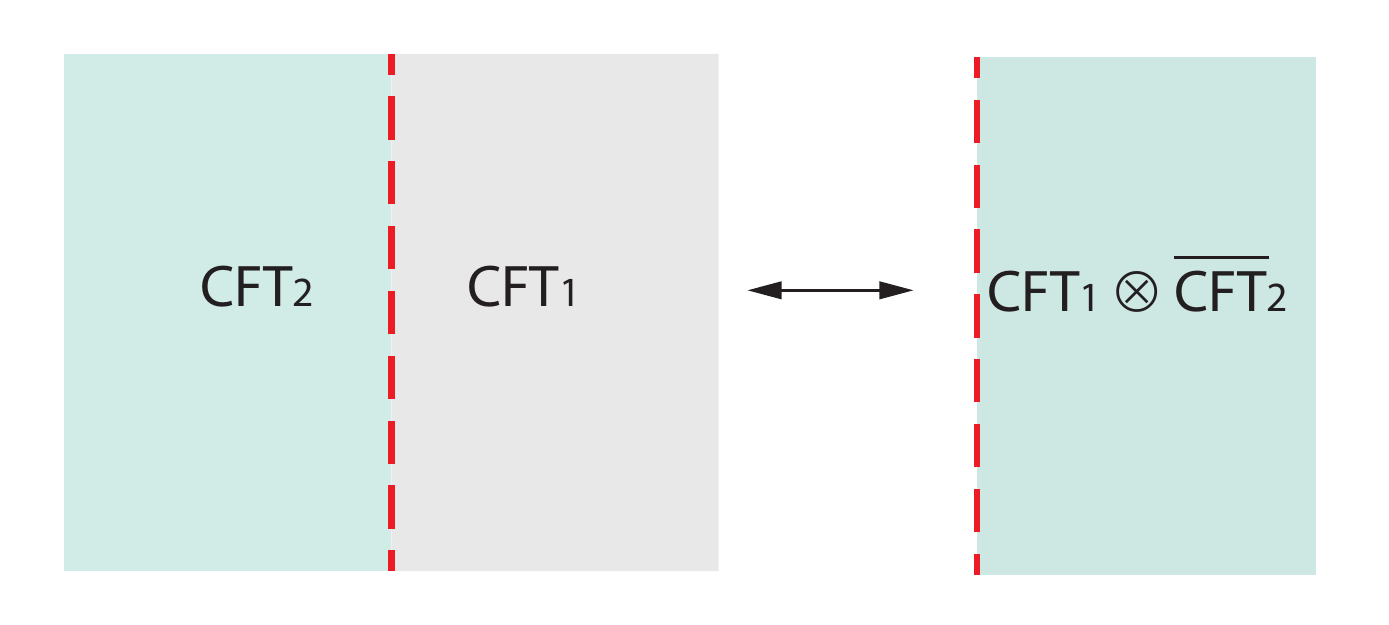}
\caption{A conformal interface (red dotted line) that connects CFT$_1$ and CFT$_2$.
Through folding and unfolding, one can relate a conformal interface
and a conformal boundary for the product theory $\text{CFT}_1\otimes \overline{\text{CFT}}_2$.
}
\label{folding}
\end{figure}

Conformal interfaces
are one dimensional objects that connect two, possibly different, conformal field theories (CFTs)
in two dimensional spacetime.\cite{Oshikawa1996,Oshikawa1997,interface01,Affleck1994,Affleck2003,Affleck2006}
A conformal interface can be described by a conformal boundary condition for the product theory
after a folding (see Fig.\ref{folding}). Considering that a conformal boundary condition preserves
the total stress tensors of the product theory,\cite{CardyBC} then one has
\be\label{T1T2}
T_1-\overline{T}_1=T_2-\overline{T}_2
\ee
along the conformal interface, where $T_i\, (\overline{T}_i)$ is the holomorphic (anti-holomorphic)
component of the stress tensor of CFT$_i$. There are two special cases for a conformal interface.
If each side of Eq.(\ref{T1T2}) equals zero, then the conformal interface itself is nothing but
a conformal boundary for CFT$_1$ and CFT$_2$ separately. That is, the two CFTs are decoupled
and this interface is totally reflective.
On the other hand, if $T_1=T_2$ and $\overline{T}_1=\overline{T}_2$, the holomorhpc/anti-holomorphic stress tensor
is continuous along the interface.
In this case, the interface commutes with the holomorphic/anti-holomorphic stress tensor and can be
freely deformed in correlators, as long as the interface does not cross any field insertion points.
The interface in the latter case is called a topological interface and is totally transmissive. \cite{Frohlich2006}
For general conformal interfaces which are partially transmissive,\cite{PartialTransmissive}
it is difficult to classify them even for the Virasoro minimal models,
since it corresponds to the classification of conformal boundary conditions
for the \textit{product} theory of Virasoro minimal models.
Nevertheless, there are some well studied examples, \textit{e.g.}, conformal interfaces in a two
dimensional Ising model,\cite{Oshikawa1996,Oshikawa1997}
and a specific one-parameter family of conformal interfaces in free boson CFTs.\cite{interface01}

In condensed matter physics, the application of conformal interfaces have been studied
in two dimensional Ising models,\cite{Oshikawa1996,Oshikawa1997} junctions of quantum wires, \cite{Affleck1994,Affleck2003,Affleck2006} etc.
 In AdS/CFT correspondence, conformal interfaces may occur when
branes extend to the boundary of the AdS-space,\cite{interface01,Karch00,Karch01,Ooguri01}
and have received extensive attention in high energy physics,
see, \textit{e.g.}, Refs.\onlinecite{interface01,Quella2002,Zuber2001,Watts2004,Fuchs2004,Bachs2004,Watts2007,Gaiotto1201,
Karch00,Karch01,Ooguri01,Gutperle1511,Gutperle1512,Gutperle1611}.

In this work, we are mainly interested in the entanglement property of CFTs in the presence of a conformal interface.
This issue was mainly studied in the ground state of two dimensional CFTs.
In Ref.\onlinecite{Sakai0809}, entanglement entropy across a conformal interface
in compact free boson CFTs was calculated.
 It was found that entanglement entropy for a subsystem $A$ has the
form
\be\label{SA_ceff}
S_A\simeq \frac{c_{\text{eff}}}{6}\log L,
\ee
where $L$ is the length of subsystem.
We use `$\simeq$' instead of `$=$' because only the leading term is concerned here.
The effective central charge $c_{\text{eff}}\in [0,c]$ depends on the
transmission property of the conformal interface (See Ref.\onlinecite{Sakai0809}
for the concrete expression of $c_{\text{eff}}$.).
In particular, $c_{\text{eff}}=0$
corresponds to a totally reflective interface,
and $c_{\text{eff}}=c=1$ corresponds to a totally transmissive interface.
Later, the generic formula in Eq.(\ref{SA_ceff}) and the expression of $c_{\text{eff}}$
were also obtained in a two dimensional Ising CFT. \cite{Brunner1505}
Most recently, quantum entanglement across a conformal interface where several CFTs join together was
studied. It was found that the entanglement entropy of a single CFT$_i$, with other CFTs as the rest,
also has the generic form in Eq.(\ref{SA_ceff}).\cite{Miller1701}

Aside from the conformal field theory approach, there are many numerical and analytical results
on tight-binding lattice models, where the entanglement entropies also show the behavior in Eq.(\ref{SA_ceff}).
In  Ref.\onlinecite{Igloi2009},
the effect of a conformal interface on entanglement entropy in the Ising
model and related fermionic systems was numerically studied.
Later in Ref.\onlinecite{Eisler1005},
analytical results of both entanglement spectrum and entanglement entropy were derived,
and the result on entanglement entropy was later confirmed in a CFT calculation.\cite{Brunner1505}
A series of works were then stimulated, including entanglement entropy across quantum wire
junctions,\cite{Calabrese1105,Calabrese1110} and entanglement entropy across a conformal interface
in bosonic quantum chains.\cite{Eisler1201}
Interestingly, in Ref.\onlinecite{Eisler1205}, entanglement evolution across a conformal interface
after a local quantum quench was analytically studied
based on a critical fermion chain.
The local quantum quench is realized by connecting two critical half-chains at their ends
through a conformal interface suddenly. Then the time evolution of entanglement
entropy for one of the half-chains is found to have the following form
\be\label{quench_ceff}
S_A(t)\simeq \frac{c'_{\text{eff}}}{3}\log t,
\ee
where $c'_{\text{eff}}\in [0,c]$ depends on the transmission property of the conformal
interface (For a totally transmissive interface, \textit{i.e.}, $c'_{\text{eff}}=c$, Eq.(\ref{quench_ceff})
 reduces to the well known result in Ref.\onlinecite{L_Quench}).
What is nontrivial, it was found that \cite{Eisler1205}
\be\label{c1_c2}
c'_{\text{eff}}=c_{\text{eff}},
\ee
where $c_{\text{eff}}$ is the effective central charge that appears in Eq.(\ref{SA_ceff}).
That is, in a critical free fermion chain, the conformal interface suppresses the entanglement entropy
in the same way for both ground-state and local-quench cases.
Then it is natural to ask the following questions:


-- The conclusion in Eq.(\ref{c1_c2}) is obtained based on a specific lattice model (a free fermion chain).
Is there a more general framework or a field theory approach to show this result?


-- What happens for other quantum quenches, \textit{e.g.},
a global quench or an inhomogeneous quench, in the presence of a conformal interface?
Does the conformal interface also suppress the entanglement evolution
in the same way as that in the ground-state case?
If so, what is the common feature/structure underlying
these different setups?


In this work, we aim to answer these questions based on the conformal field theory approach.
The main results are as follows.

\begin{enumerate}

\item

We show that the result $c'_{\text{eff}}=c_{\text{eff}}$ for a local quench in Eq.\,\eqref{c1_c2}
can be obtained by using conformal mappings.
Compared to the previous work which focuses on a critical lattice model,\cite{Eisler1205}
our approach is universal and applies to arbitrary CFTs.

\item

We generalize the result in Eq.\,\eqref{c1_c2} to other quantum quenches including
a global quench and a specific inhomogeneous quench.
It is found that the setups for the ground-state case and the quantum-quench cases
can be conformally mapped to the same configuration.
This is why the entanglement entropy is suppressed by a conformal interface in the same way
for all these cases.

\item

We confirm our CFT results with numerical calculations based on a critical lattice model.
Furthermore, based on the quasi-particle picture, we conjecture that
for an arbitrary quantum quench with the initial state described by a regularized
conformal boundary state, the conformal interface suppresses the time evolution
of entanglement entropy by effectively replacing the central charge $c$ with $c_{\text{eff}}$,
where $c_{\text{eff}}$ is the same as that in the ground-state case in Eq.(\ref{SA_ceff}).

\end{enumerate}

The structure of this work is organized as follows. In Sec.\ref{Sec_EE_Evolution},
we study how a conformal interface suppresses the time evolution of entanglement entropy after
quantum quenches including global, local, and a certain inhomogeneous quenches.
Then in Sec.\ref{Sec_Lattice}, we confirm our field-theory results
based on a lattice model calculation. In Sec.\ref{Sec_Quasiparticle}, we discuss the quasi-particle picture and
its application in quantum quenches with a conformal interface,
including the case of a global quench with a conformal interface inside the subsystem,
and the case of an arbitrary quench with a conformal interface along the entanglement cut.
Then we conclude our work in Sec.\ref{Sec_conclusion}.
There are two appendices on the effect of conformal boundary
conditions on the entanglement entropy, and the left-right entanglement of a conformal boundary state
as entanglement sources of real-space entanglement after a quantum quench.

\section{Evolution of entanglement entropy across a conformal interface}
\label{Sec_EE_Evolution}

\begin{figure}[htp]
\includegraphics[width=3.20in]{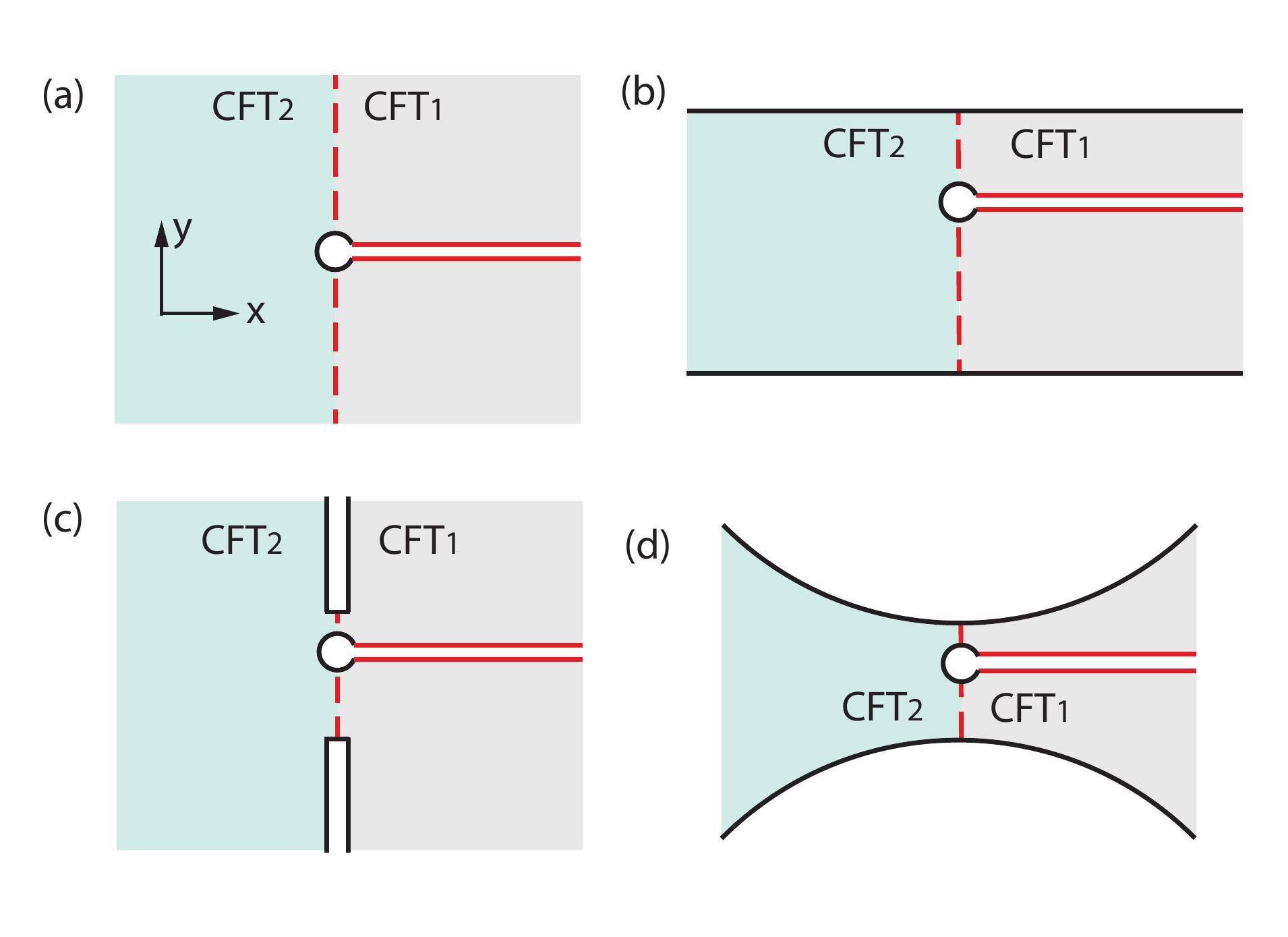}
\caption{Setups for
two CFTs with a conformal interface in the case of (a) a ground state, (b) a global quench, (c) a local quench and (d) an
inhomogeneous quench.
The red dotted line along $x=0$ represents the conformal interface connecting $\text{CFT}_1$
and $\text{CFT}_2$, and the red solid line is the branch cut in subsystem $A$. We remove a small disc
at the entangling point $z=0+i\tau$ as a UV regularization.
}\label{Interface}
\end{figure}

Now we are mainly interested in the time evolution of entanglement entropy
after a quantum quench in the presence of a conformal interface.
In particular, we will study a global quench\cite{G_Quench1, G_Quench2}, a local quench\cite{L_Quench} and
an exactly solvable inhomogeneous quench\cite{Inhomo_Quench} as examples.
Throughout this work, unless explicitly interpreted,
the conformal interface we consider is along the entanglement cut $x=0$.

Our basic strategy is as follows.
Through conformal mappings, we show that for both the ground-state case and the non-equilibrium cases, the
path integral representation of the reduced density matrix can be mapped to the same configuration.
The conformal mappings we use are well explored in the recent work by Cardy and Tonni,\cite{Cardy1608}
and the new ingredient in this work is to include conformal interfaces, and see how these conformal interfaces
transform after conformal mappings.

Before studying quantum quenches, it is helpful to have a short review of the
ground-state case with a conformal interface. \cite{Sakai0809,Brunner1505}

\subsection{Brief review of entanglement entropy in ground state
with a conformal interface}

\begin{figure}[t]
\includegraphics[width=2.70in]{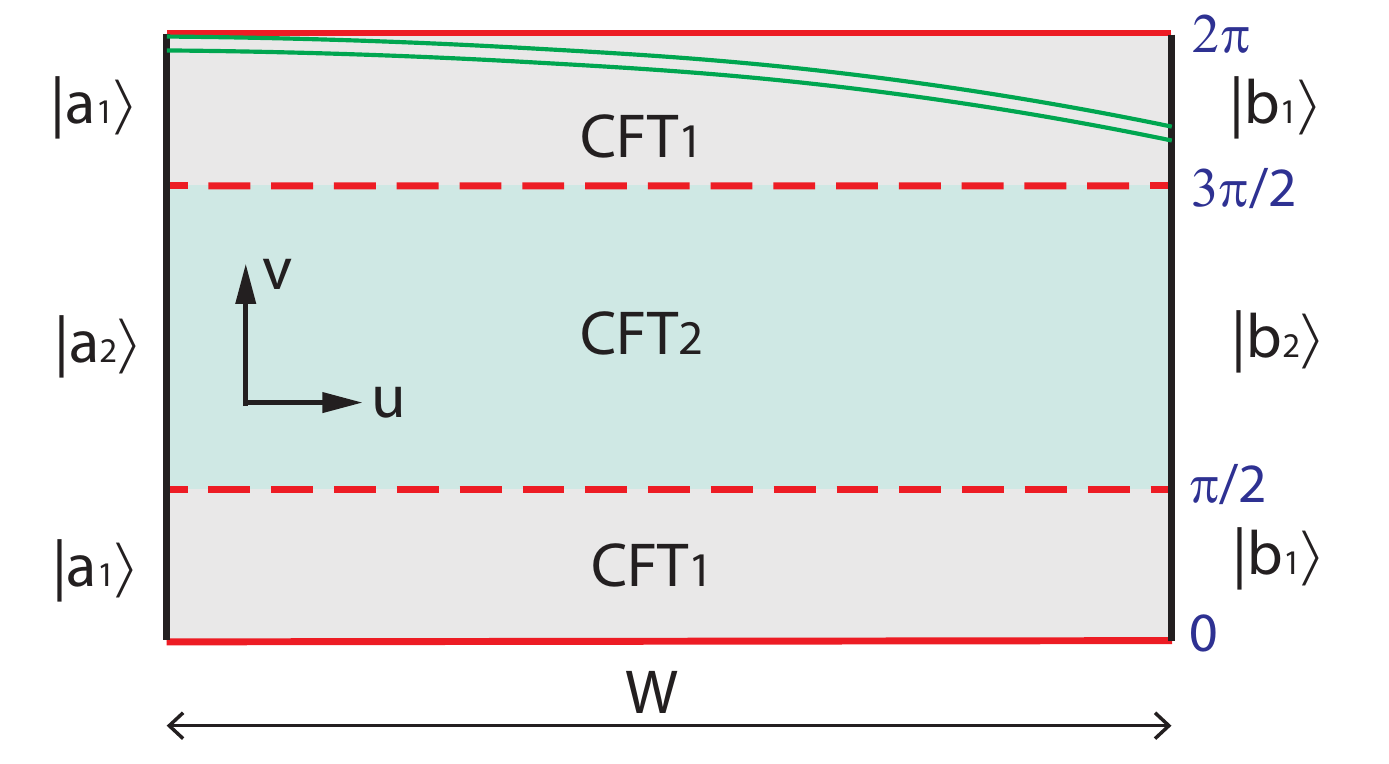}
\caption{
After conformal mappings, different setups for $\rho_A$ in Fig.\ref{Interface} are mapped to the same
configuration in $w$-plane, with $w=u+iv$.
Now the conformal interface (dotted line) is along $v=\pi/2$ and $3\pi/2$.
The difference is the location of branch cuts.
For equilibrium case in Fig.\ref{Interface} (a), the branch cuts are mapped to
$v=0$ and $v=2\pi$. For non-equilibrium cases such as the global quench in Fig.\ref{Interface} (b),
the branch cuts are mapped to the green solid lines, which are no longer along constant $v$.
}\label{AfterMap}
\end{figure}

Now we have two CFTs connected by a conformal interface along the
entanglement cut $x=0$ [see Fig.\ref{Interface} (a)] .
Subsystem $A$ corresponds to CFT$_1$ in the right half plane.
Then one introduces a UV cutoff $|z|=\epsilon$ and an IR cutoff $|z|=L$,
and imposes conformal boundary conditions $|a_1\rangle$, $|a_2\rangle$ along $|z|=\epsilon$, and $|b_1\rangle$, $|b_2\rangle$
along $|z|=L$, respectively.\cite{RemarkImposeBC}
The reduced density matrix $\rho_A=\text{Tr}_{\bar{A}}\rho$ could be viewed as
the partition function defined on this manifold, with the
branch cuts along $\{x+iy|\epsilon\le x\le L, y=0^{\pm}\}$.
Then we consider the following conformal mapping
\be
w=\log z,
\ee
which maps the configuration in Fig.\ref{Interface} (a) to a strip in $w$-plane in Fig.\ref{AfterMap}.
One can find that the conformal interfaces are mapped to two straight lines along
$v=\text{Im}\,w=\frac{\pi}{2},\frac{3\pi}{2}$, the
branch cuts are mapped to straight lines along $v=0,2\pi$, and the boundaries along $|z|=\epsilon$ and $|z|=L$
 are mapped to $u=\text{Re}\,w=\log\epsilon$ and $\log L$, respectively.
It is noted that the conformal boundary conditions $|a_1\rangle$ and $|a_2\rangle$ in $w$-plane are along $u=\log \epsilon$ with
$v=(0,\frac{\pi}{2})\cup (\frac{3\pi}{2},2\pi)$ and $v=(\frac{\pi}{2},\frac{3\pi}{2})$, respectively.
(See also Fig.\ref{AfterMap} for $|b_1\rangle$ and $b_2\rangle$ along $u=\log L$.)
The width of strip in $\text{Re}\,w$ direction is
\be\label{W_ground}
W=\log L-\log \epsilon=\log \frac{L}{\epsilon}.
\ee
Based on the configuration in Fig.\ref{AfterMap}, one can express the Renyi entropy as
\be\label{Renyi_Zn}
S_A^{(n)}:=\frac{1}{1-n}\log \frac{\text{Tr}\left(\rho_A^n\right)}{\left(\text{Tr}\,\rho_A\right)^n}=\frac{1}{1-n}\log \frac{Z_n}{(Z_1)^n},
\ee
where $Z_1=\text{Tr}\,\rho_A$ is the partition function on a cylinder, which is obtained by gluing
the two branch cuts along $v=0$ and $v=2\pi$ in Fig.\ref{AfterMap}.
Similarly, by considering $n$ copies of strips in Fig.\ref{AfterMap},
and gluing the branch cuts one by one, we can obtain the partition function $Z_n$.
To evaluate the partition function $Z_n$ with $2n$ conformal interfaces intersecting with two boundaries
is not an easy task.
Ref.\onlinecite{Sakai0809} and following works\cite{Brunner1505} circumvent this difficulty
by taking periodic boundary conditions along $u=\text{Re}\,w$ direction, so that the cylinder becomes a torus.

A remark here:
It is noted that changing the open boundary condition of a cylinder with $W\gg 1$ in Fig.\ref{AfterMap}
to a periodic boundary condition does not affect the leading term of entanglement entropy,
as discussed in Appendix.\ref{Effect_BC}. In this work, since we are mainly interested in
the leading term of entanglement entropy, hereafter we will take periodic boundary
condition in $\text{Re}\,w$ direction in Fig.\ref{AfterMap}.

Then the partition function $Z_n$ can be expressed as
\be\label{PartitionZn}
Z_n=\text{Tr}\left(I_{12}q^{H_{\text{CFT}_2}}I_{12}^{\dag}q^{ H_{\text{CFT}_1}}\right)^n,
\ee
where $H_{\text{CFT}_{1,2}}$ is the hamiltonian operator in the respective CFT,
$I_{12}$ denotes the interface operator,\cite{InterfaceOperator}
and $q=e^{-\pi \delta}$ with $\delta=2\pi/W$.
The partition function $Z_n$ and entanglement entropy $S_A$ has been explicitly evaluated for free
boson and free fermion CFTs,\cite{Sakai0809,Brunner1505}.
It was found that
\be\label{SAground}
S_A=\lim_{n\to1}S_A^{(n)} \simeq \frac{c_{\text{eff}}}{6}W,
\ee
where $c_{\text{eff}}\in [0,c]$ depends on the transmission property of the conformal interface,
as well as the type of CFTs.
Then based on Eqs.(\ref{W_ground}) and (\ref{SAground}), one can obtain
the entanglement entropy of subsystem $A$ as follows
\be\label{SAground2}
S_A\simeq\frac{c_{\text{eff}}}{6}\log\frac{L}{\epsilon}.
\ee

Here we give two remarks on Eq.\eqref{SAground}.

-- Remark 1: 

Here $c_{\text{eff}}\in [0,c]$ refers to the case that CFT$_1$ and CFT$_2$ have the same central charge, 
\textit{i.e.}, $c_1=c_2=c$.
In the case $c_1\neq c_2$, the effective central charge satisfies
\be
c_{\text{eff}}\in \big[
0, \text{min}[c_1,c_2]
\big].
\ee
This is because the degrees of freedom that contribute to the coupling of two CFTs at the conformal interface are
upper bounded by $\text{min}[c_1,c_2]$.

-- Remark 2:

It is noted that the formula in Eq.\eqref{SAground} was explicitly derived in
free boson and free fermion CFTs. Given a generic CFT, since the classification of a
conformal interface is not well understood (see the introduction),
an explicit calculation of entanglement entropy with a conformal interface is still absent.
Nevertheless, based on the following argument, one can find that 
Eq.\eqref{SAground} is valid for generic CFTs.

For simplicity, let us consider the case CFT$_1$=CFT$_2$, with the reduced density matrix
represented in Fig.\ref{AfterMap} after an appropriate conformal mapping.
We evaluate the partition function $Z_n$ [see Eq.\eqref{Renyi_Zn}] as follows:
\be
Z_n=\text{Tr} e^{-H_n W},
\ee
where $H_n$ is the Hamiltonian defined on a circle of circumference $2n\pi$, 
with $2n$ conformal interfaces inserted (see Fig.\ref{AfterMap} for $n=1$).
Since $W\gg 1$, the ground state energy of $H_n$ dominates in $Z_n$, \textit{i.e.},
\be
Z_n\simeq e^{-E_n\, W}.
\ee 
We emphasize that the ground state energy $E_n$ depends on the transmission property of the conformal interface.
Then based on Eq.\eqref{Renyi_Zn}, one can find the leading term of the entanglement entropy has the form
\be
S_A^{(n)}=\frac{nE_1-E_n}{1-n} W=: \frac{c_{\text{eff}}}{12}\cdot \frac{1+n}{n}W,
\ee
where we have defined the effective central charge as
\be\label{c_eff_Energy}
c_{\text{eff}}:=\frac{12 n}{1-n^2}\cdot (nE_1-E_n).
\ee
That is, the effective central charge can be defined through the ground state energy in the presence of conformal interfaces.
For the case that the conformal interface is transparent, one has $E_n=-\frac{c}{12 n}$ and $E_1=-\frac{c}{12}$, and then
one can find $c_{\text{eff}}=c$, as expected. For the case that the conformal interface is totally reflective, one has
$E_n=-\frac{nc}{12}$ and $E_1=-\frac{c}{12}$, and then $c_{\text{eff}}=0$, also as expected.

For a conformal interface with partial transmission and partial reflection, $c_{\text{eff}}$ in Eq.\eqref{c_eff_Energy}
is not easy to evaluate, and only the free bosons and free fermions have been studied.
Nevertheless, we have shown that the formula in Eq.\eqref{SAground} is valid for generic CFTs.

As a final remark, it is not straightforward to see that $c_{\text{eff}}$ in Eq.\eqref{c_eff_Energy} is smaller than $c$.
We believe that $c_{\text{eff}}\le c$ is true, because a conformal interface with partial transmission reduces the degrees of
freedom coupling to the other CFT, and therefore reduces the entanglement entropy.

\subsection{Quantum quenches
with a conformal interface}
\label{QuenchCases}

In the following parts, we will show that different setups of quantum quenches,
after conformal mappings $w=f(z)$,
may be mapped to the same configuration in Fig.\ref{AfterMap}.
In particular, the partition function $Z_n=\text{Tr}\left(\rho_A^n\right)$
is exactly the same as Eq.(\ref{PartitionZn}) up to the concrete form of $W$, which is the width
of strip (see Fig.\ref{AfterMap}).
By repeating the same calculation in Refs.\onlinecite{Sakai0809, Brunner1505},
one can obtain the entanglement entropy
\be\label{S_nonEquilibrium}
S_A(t)\simeq \frac{c_{\text{eff}}}{6}W(t),
\ee
where the effective central charge $c_{\text{eff}}$
is exactly the same as that in the ground state in Eqs.(\ref{SAground}) and (\ref{SAground2}).

\subsubsection{Global quench}
\label{subsubSec: GlobalQuench}

We follow Refs.\onlinecite{G_Quench1, G_Quench2} for the setup of a global quench.
One starts from a short-range entangled initial state $|\phi_0\rangle$,
and has it evolve with a CFT hamiltonian $H_{\text{CFT}}$.
For simplicity, one may choose $|\phi_0\rangle$ in the following form
\be\label{InitialState}
|\phi_0\rangle=e^{-\beta H_{\text{CFT}}/4}|b\rangle,
\ee
where $|b\rangle$ is the so-called conformal boundary state, and has no real space entanglement. \cite{Miyaji1412}
By evolving $|b\rangle$ with an imaginary time $\beta/4$, the initial state $|\phi_0\rangle$ has finite
real-space entanglement and is normalizable.
The reason we choose the factor $\beta/4$ instead of $\beta$ is because the expectation value
$\langle \phi_0|H_{\text{CFT}}|\phi_0\rangle$ is the same as that in finite temperature $1/\beta$.
Then the time dependent density matrix has the form
$\rho(t)=e^{-iH_{\text{CFT}}t}e^{-\beta H_{\text{CFT}}/4}|b\rangle\langle b|e^{-\beta H_{\text{CFT}}/4}e^{iH_{\text{CFT}}t}$, which in Euclidean spacetime becomes
$
\rho(\tau)=e^{-H_{\text{CFT}}\tau}e^{-\beta H_{\text{CFT}}/4}|b\rangle
$
$\langle b|e^{-\beta H_{\text{CFT}}/4}e^{+H_{\text{CFT}}\tau}.
$

Now, for two CFTs connected by a conformal interface, in general we need to modify the initial state
by taking $|b\rangle=|b_1\rangle$ for $x>0$, and $|b\rangle=|b_2\rangle$ for $x<0$, where $|b_{1(2)}\rangle$
is the conformal boundary state corresponding to CFT$_{1(2)}$, as shown in Fig.\ref{Interface} (b).
This is similar to how we impose boundary conditions
along $|z|=\epsilon$ and $|z|=L$ in the ground-state case.
It is noted that this kind of ``domain-wall" initial state
without a conformal interface
 has been studied in Ref.\onlinecite{DomainWall0804}. It was found
that only the constant term in entanglement entropy is modified.
Here in our setup the ``domain-wall" initial state is very natural because we may consider two different CFTs.
Similar to the ground-state case,
we remove a small disc with radius $\epsilon$ around $z=0+i\tau$ as a UV regularization.
Then the path integral representation of the reduced density matrix $\rho_A=\text{Tr}_{\bar{A}}\,\rho$
is shown in Fig.\ref{Interface} (b).

Now we consider the following conformal mapping\cite{Cardy1608}
\be
w=f(z)=\log \left(
\frac
{\sinh[\pi(z-i\tau)/\beta]}
{\cosh[\pi(z+i\tau)/\beta]}
\right),
\ee
which maps the configuration in Fig.\ref{Interface} (b) to the strip in $w$-plane in Fig.\ref{AfterMap}.
In particular, one can find that the conformal interface along $x=0$ in $z$-plane is mapped to
\be
w(0+iy)=\left\{
\begin{split}
&i\frac{\pi}{2}+ \log \left(\frac{\sin \frac{\pi}{\beta}(y-\tau)}{\cos \frac{\pi}{\beta}(y+\tau)}\right),
\tau<y<\frac{\beta}{4},\\
&i\frac{3\pi}{2}+\log \left(\frac{\sin \frac{\pi}{\beta}(\tau-y)}{\cos \frac{\pi}{\beta}(y+\tau)}\right),
-\frac{\beta}{4}<y<\tau.\\
\end{split}
\nonumber
\right.
\ee
That is, the conformal interfaces are mapped to straight lines along $\text{Im}\, w=\frac{\pi}{2}$ and $\frac{3\pi}{2}$ in the strip
in $w$-plane.

Different from the equilibrium case, now the branch cuts (solid red lines in Fig.\ref{Interface} (b))
are mapped to curves (green solid lines in Fig.\ref{AfterMap})
which are no longer along $v=0,2\pi$.\cite{Cardy1608} However, we emphasize that the location of
branch cuts has no effect on the partition function
$Z_n=\text{Tr}\left(\rho_A^n\right)$, in which the branch cuts are glued one by one.
Therefore, hereafter we will no longer show explicitly the location of branch cuts
for local and inhomogeneous quenches.

Considering the UV cutoff $|z|=\epsilon$ at the entangling point $z=0+i\tau$, then the width of strip
in Fig.\ref{AfterMap} can be evaluated as $W=\big|w(0+i(\tau+\epsilon))-w(0+i\frac{\beta}{4})\big|=
\big|w(0+i(\tau-\epsilon))-w(0-i\frac{\beta}{4})\big|\simeq \log\left(\frac{\cos(2\pi\tau/\beta)}{\sin(\pi\epsilon/\beta)}\right)$.
After analytical continuation $\tau\to it$, one has
\be
W(t)\simeq \log \left(
\frac{\cosh(2\pi t/\beta)}
{\sin(\pi\epsilon/\beta)}
\right)\simeq \frac{2\pi t}{\beta}+\log\frac{\beta}{2\pi\epsilon}.
\ee
Then based on Eq.(\ref{S_nonEquilibrium}), one can obtain
the leading term of entanglement entropy as
\be\label{SAt_Global}
S_A(t)\simeq \frac{c_{\text{eff}}}{6}W(t)\simeq \frac{\pi c_{\text{eff}}}{3\beta}t.
\ee
For a totally transmissive conformal interface, \textit{i.e.}, $c_{\text{eff}}=c$,
one recovers the well known result $S_A(t)\simeq \frac{\pi c}{3\beta}t$.\cite{G_Quench1, G_Quench2}

\subsubsection{Local quench}

For local quantum quenches in a CFT, there are mainly two interesting setups:
one is the ``cut-and-glue" setup, which connects two CFTs at their ends suddenly,\cite{L_Quench}
and the other is to act on the CFT with a local operator.\cite{Nozaki1401,Nozaki2014,He1403,Caputa2014,Nozaki2014,Caputa2015,He1501}

Here we are interested in the ``cut-and-glue" setup.
That is, before we glue the two CFTs at $t=0$, each CFT stays in its ground state.
Then at $t=0$, the two CFTs are connected through a conformal interface.
Following Ref.\onlinecite{L_Quench}, in Euclidean spacetime, one may consider two slits:
one slit goes from $z=0+i\infty$ to $z=0+i\lambda$, and the other slit goes from
$z=0-i\infty$ to $z=0-i\lambda$, as shown in Fig.\ref{Interface} (c).
Along the slits, conformal boundary condition $|b_1\rangle$ ($|b_2\rangle$)
is imposed on the CFT$_1$ (CFT$_2$) side.
As before, we remove a small disc of radius $\epsilon$ at $z=0+i\tau$, and then the remaining part
can be mapped to a strip in $w$-plane in Fig.\ref{AfterMap} by considering the following conformal
mapping \cite{Cardy1608}
\be
w=\log
\left(
\frac{
\sqrt{(\lambda^2-\tau^2)(z^2+\lambda^2)}-i\tau z-\lambda^2
}
{\lambda(z-i\tau)}
\right).
\ee
It is straightforward to check that the conformal interface along $x=0$ is mapped to
\begin{small}
\be
w(0+iy)=i\frac{\pi}{2}
+
\log\left(
\frac
{
(\lambda^2-\tau y)-\sqrt{(\lambda^2-\tau^2)(\lambda^2-y^2)}
}
{\lambda(y-\tau)}
\right),
\nonumber
\ee
\end{small}
for $\tau<y<\lambda$, and
\begin{small}
\be
w(0+iy)=i\frac{3\pi}{2}
+
\log\left(
\frac
{
(\lambda^2-\tau y)-\sqrt{(\lambda^2-\tau^2)(\lambda^2-y^2)}
}
{\lambda(\tau-y)}
\right),
\nonumber
\ee
\end{small}
for $-\lambda<y<\tau$. That is, the conformal interface along $x=0$ in $z$-plane
is mapped to two straight lines along $\text{Im}\,w=\frac{\pi}{2},\frac{3\pi}{2}$ in $w$-plane.
One can check that the width of the strip in Fig.\ref{AfterMap} is
$W=|w(0+i\lambda)-w(0+i(\tau+\epsilon))|=
|w(0-i\lambda)-w(0+i(\tau-\epsilon))|
\simeq \log \frac{2(\lambda^2-\tau^2)}{\lambda\epsilon}$.
After analytical continuation $\tau\to it$, one has
\be
W(t)\simeq \log\frac{2(\lambda^2+t^2)}{\epsilon \lambda}.
\ee
Therefore, based on Eq.(\ref{S_nonEquilibrium}), one can obtain
the leading term of entanglement entropy $S_A(t)$ as follows
\be
S_A(t)\simeq \frac{c_{\text{eff}}}{6}W(t)\simeq \frac{c_{\text{eff}}}{3}\log t,
\ee
which agrees with the lattice model results in Eqs.(\ref{quench_ceff}) and (\ref{c1_c2}).
It also recovers the well known result $S_A(t)\simeq \frac{c}{3}\log t$ for a totally transmissive
conformal interface, as expected.

\subsubsection{Inhomogeneous quench}
\label{Sec_Inhomogeneous_Quench}

Among different setups of inhomogeneous quantum quenches\cite{Sot0808,Calabrese0804,Viti1507,Allegra1512}
(see also Ref.\onlinecite{Quench_Review} for a review),
here we are mainly interested in the case with smoothly varying initial state.\cite{Sot0808}
Then $\beta$ in the initial state $|\phi_0\rangle$ in Eq.(\ref{InitialState}) is no longer a constant,
but depends on the position $x$. Then $|\phi_0\rangle$ can be explicitly written as
\be\label{phi0_inhom}
|\phi_0\rangle=e^{-\frac{1}{4}\int\beta(x)\mathcal{H}_{\text{CFT}}(x)dx}|b\rangle,
\ee
where $\mathcal{H}_{\text{CFT}}(x)$ is the hamiltonian density.
Here we consider a solvable inhomogeneous quantum quench, with
$\beta(x)$ chosen as follows\cite{Inhomo_Quench}
\be\label{Inhomo_quench}
\beta(x)=4\sin\beta_0\cdot \sqrt{\Lambda^2+\left(\frac{x}{\cos\beta_0}\right)^2},
\ee
where $\beta_0\ll 1$ is a positive constant.
This kind of inhomogeneous quench is interesting since it shows features of a global quench in the short time limit ($t\ll \Lambda$)
and a local quench in the long time limit ($t\gg\Lambda$).\cite{Inhomo_Quench}
Now we introduce a conformal interface which is defined along $x=0$ with
$-\Lambda\sin\beta_0\le y\le \Lambda\sin\beta_0$.
The path integral representation of the the reduced density matrix $\rho_A$
is shown in  Fig.\ref{Interface} (d).

To map the configuration in Fig.\ref{Interface} (d) to
the strip in $w$-plane (see Fig.\ref{AfterMap}), we consider
the following conformal mapping
\be
w=\log \left[\frac{1+\bar{\xi}_0}{1+\xi_0}\cdot \frac{\xi(z)-\xi_0}{\xi(z)+\bar{\xi}_0}\right],
\ee
where
\be
\xi(z)=\exp\left[\frac{\pi}{2\beta_0}\sinh^{-1}\left(\frac{z}{\Lambda}\right)\right].
\ee
Here the effect of $\xi(z)$ is to map the configuration Fig.\ref{Interface} (d) to a right half plane.
By denoting $\alpha'=\frac{\pi}{2\beta_0}\arcsin \frac{y}{\Lambda}$,
and $\alpha =\frac{\pi}{2\beta_0}\arcsin \frac{\tau}{\Lambda}$,
one can find that the conformal interface in Fig.\ref{Interface} (d)  is mapped to
\be
w(0+iy)=
i\frac{\pi}{2}+ \log \left(\frac{\sin\alpha'-\sin\alpha}{1+\cos(\alpha+\alpha')
}\right),
\ee
for $\tau<y<\Lambda\sin\beta_0$, and
\be
w(0+iy)=
i\frac{3\pi}{2}+\log \left(\frac{
\sin\alpha-\sin\alpha'
}{
1+\cos(\alpha+\alpha')
}\right),
\ee
for $-\Lambda\sin\beta_0<y<\tau$.
That is, the conformal interface along $x=0$ in $z$-plane in Fig.\ref{Interface} (d) is indeed mapped to the
straight lines along $\text{Im}\,w=\frac{\pi}{2}, \frac{3\pi}{2}$ in $w$-plane in Fig.\ref{AfterMap}.
Now let us check the width $W$ of the strip in Fig.\ref{AfterMap}, which may be expressed as
$
W=\big|w(i\Lambda\sin\beta_0)-w[i(\tau+\epsilon)]\big|=\big|w[i(\tau-\epsilon)]-w(-i\Lambda\sin\beta_0)\big|.
$
After some simple algebra, one can find that
$
W(t)\simeq \frac{\pi}{2\beta_0}\log \left(
\sqrt{1+\frac{t^2}{\Lambda^2}}+\frac{t}{\Lambda}
\right).
$
Then, according to Eq.(\ref{S_nonEquilibrium}), the time evolution of entanglement entropy
of subsystem $A$ (or CFT$_1$) has the form
\be\label{SA_inhomogeneous}
S_A(t)\simeq \frac{\pi c_{\text{eff}}}{12\beta_0}\log \left(
\sqrt{1+\frac{t^2}{\Lambda^2}}+\frac{t}{\Lambda}
\right),
\ee
which shows interesting limits
\be
S_A(t)\simeq
\left\{
\begin{split}
&\frac{\pi c_{\text{eff}}}{12\beta_0\Lambda}\cdot t, \quad t\ll \Lambda\\
&\frac{c_{\text{eff}}}{3}\cdot \frac{\pi}{4\beta_0}\cdot \log t, \quad t\gg \lambda.
\end{split}
\right.
\ee
Comparing with the result in Ref.\onlinecite{Inhomo_Quench}, here the conformal interface
suppresses the entanglement evolution $S_A(t)$ by replacing $c$ with $c_{\text{eff}}$.

\section{Lattice calculation}
\label{Sec_Lattice}

In this section we will check our field theory results based on a critical lattice model.
Critical lattice models with a conformal interface have been studied in
different cases, including a critical Ising model,\cite{interface01,Oshikawa1996} a harmonic chain,\cite{Eisler1201}
and a free fermion chain.\cite{Eisler1005,Eisler1201,Eisler1205}
Here,
we will take a critical free fermion chain for example.

We study a free fermion chain of length $2L$, with a conformal interface located
between sites $L$ and $L+1$.
The hamiltonian has the following expression:\cite{Eisler1205}
\be\label{H_ConformalI}
H=\frac{1}{2}\sum_{i,j=1}^{2L}H_{i,j}c_i^{\dag}c_j,
\ee
where the nonzero elements are
\be\label{H_Conformal01}
H_{i,i+1}=H_{i+1,i}=
\left\{
\begin{split}
&-1,\quad i\neq L,\\
&-\lambda,\quad i=L,
\end{split}
\right.
\ee
and
\be\label{H_Conformal02}
H_{L,L}=-H_{L+1,L+1}=\sqrt{1-\lambda^2}.
\ee
Apparently, for $\lambda=1$, the fermion chain is homogeneous, and there is no interface/defect;
for $\lambda=0$, we have two decoupled chains.
Curious readers may wonder why we choose the interface of the form
in Eqs.(\ref{H_Conformal01}) and (\ref{H_Conformal02}).
In Ref.\onlinecite{Eisler1205}, it was found that the transmission coefficient for
this kind of interface is independent of the wavelength of incoming waves, and thus is scale invariant.
On the other hand, if we simply choose a bond defect in Eq.(\ref{H_Conformal01}), one can find that
the transmission coefficient is wavelength dependent, and is non-conformal.

In the following, we study how the conformal interface suppresses
the entanglement entropy by extracting $c_{\text{eff}}(\lambda)/c(\lambda=1)$ for different cases.
In particular, different kinds of quantum quenches are realized through different initial states, which evolve
according to the same hamiltonian in Eqs.(\ref{H_ConformalI})$\sim$(\ref{H_Conformal02}).

\textit{(a) Ground state}

\begin{figure}[t]
\includegraphics[width=3.7in]{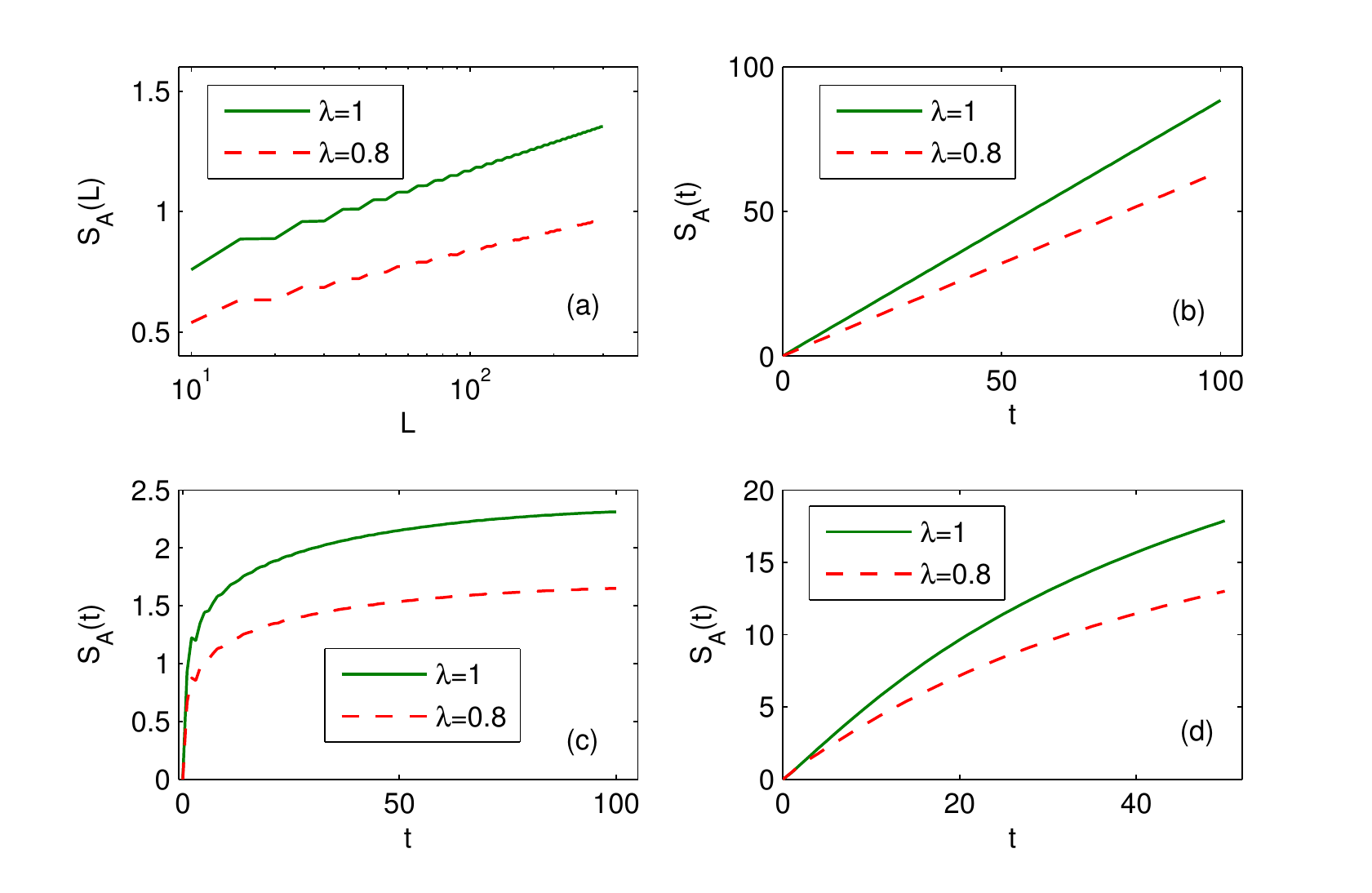}
\caption{
Numerical results for the entanglement entropy $S_A$ in different cases: (a) in the ground state, (b) after a global quench,
(c) after a local quench, and (d) after an inhomogeneous quench.
In each plot, we show the results with ($\lambda=0.8$)
and without a conformal interface ($\lambda=1$).
}\label{4plot}
\end{figure}

We consider a free fermion chain of length $2L$ with the hamiltonian in Eq.(\ref{H_ConformalI}),
and prepare the system in the ground state.
Then we study the entanglement entropy $S_A$ for subsystem $A=[L+1,2L]$.
For $\lambda=1$, the entanglement entropy depends on $L$ as $S_A(L)=\frac{c}{6}\log L+\mathrm{const}.$, with
$c=1$ here; for
$0\le \lambda<1$, the introduction of conformal interface will suppress the entanglement entropy as
$S_A(L)=\frac{c_{\text{eff}}}{6}\log L+\mathrm{const.}$, with $0\le c_{\text{eff}}<c$.
It is noted that here the subleading constant term in $S_A$ usually depends on the parameter $\lambda$.\cite{Sakai0809,RelativeE}
By changing the length $L$, we can extract the effective central charge $c_{\text{eff}}(\lambda)$ by fitting
the numerical plot. (See Fig.\ref{4plot} (a) for a typical plot of $S_A(L)$ with $\lambda=0.8$ and $\lambda=1$.)

\textit{(b) Global quench}

\begin{figure}[t]
\includegraphics[width=3.0in]{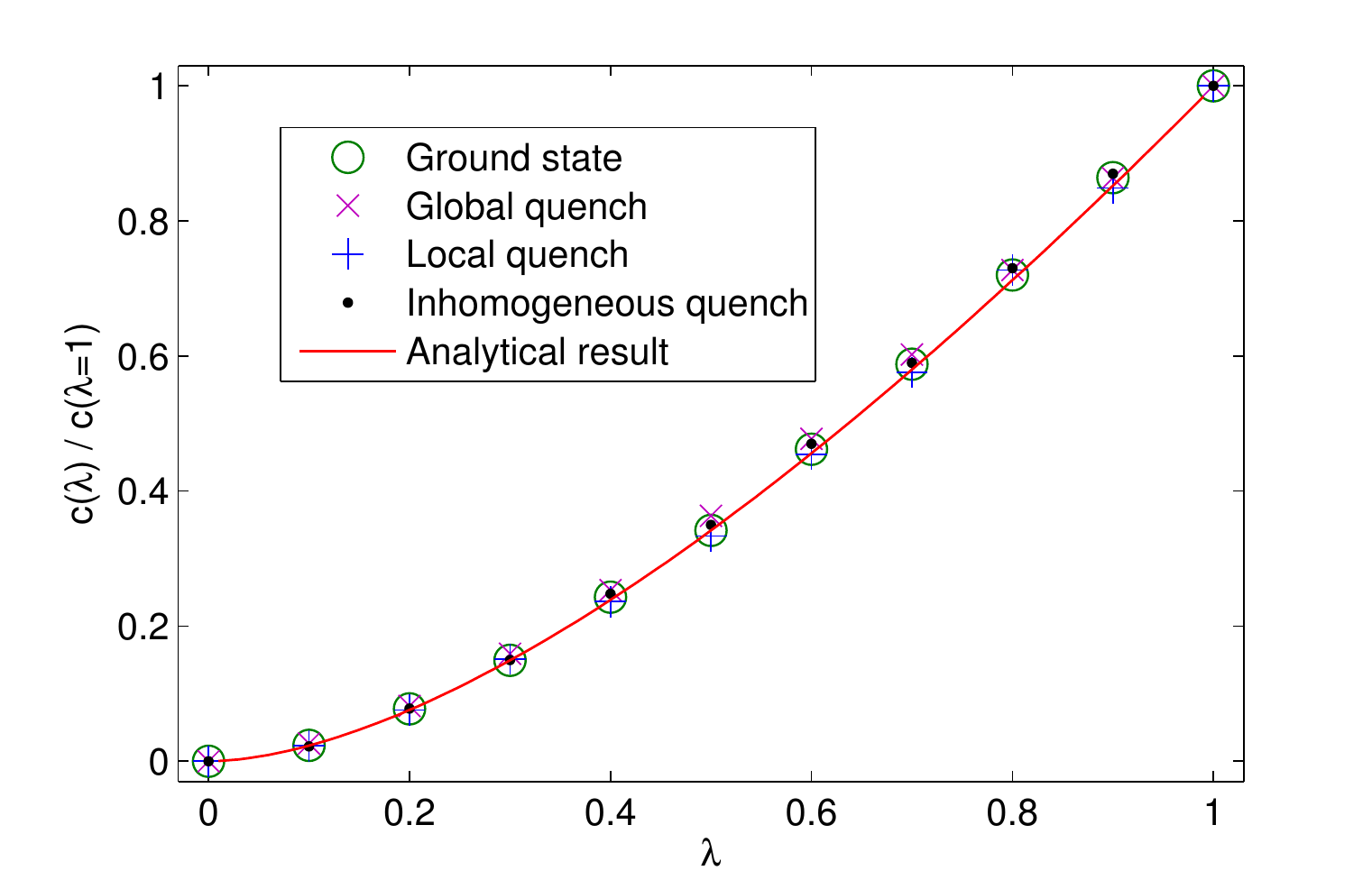}
\caption{
Numerical results for $c_{\text{eff}}(\lambda)/c(\lambda=1)$ as
a function of $\lambda$ for different cases, and the analytical result
in Eq.(\ref{AnalyticalCeff}).
$\lambda=1$ corresponds to the case with no conformal interface, and $\lambda=0$ corresponds to
the case where a critical chain is decoupled into two halves. For the free fermion chain considered here, one has
$c(\lambda=1)=1$.
}\label{Compare}
\end{figure}

We prepare the initial state $|\psi_0\rangle$ as the ground state of a massive fermion chain, by adding a mass term
to the Hamiltonian in Eq.(\ref{H_ConformalI}) [See also Ref.\onlinecite{Miyaji1412} for details.].
Then from $t=0$, we have the state evolve according to the hamiltonian in Eq.(\ref{H_ConformalI}),
and observe how the entanglement entropy $S_A(t)$ evolves, where $A=[L+1,2L]$. In the presence of a conformal interface,
it is observed that $S_A(t)=\frac{\pi c_{\text{eff}}}{3\beta}t+\mathrm{const.}$ (see Fig.\ref{4plot}
and Eq.\eqref{SAt_Global} ), where
$\beta$ is a non-universal constant and depends on the mass term in the massive fermion chain 
(See Appendix.\ref{Sec_GlobalQuench_MassTerm} for a detailed discussion.).
In the fitting procedure, for different $\lambda$, we fix the parameter $\beta$ and $L$.
Then we can obtain $c_{\text{eff}}(\lambda)/c(\lambda=1)$ in Fig.\ref{Compare}.

\textit{(c) Local quench}

This case was studied in Ref.\onlinecite{Eisler1205}. Here we briefly review the procedures to extract $c_{\text{eff}}(\lambda)$.
We prepare the initial state $|\psi_0\rangle$ as the ground state of two decoupled critical fermion chains of length $L$.
Then at $t=0$, we connect the two decoupled chains with a conformal interface.
That is, the initial state $|\psi_0\rangle$ evolves according to the hamiltonian in Eq.(\ref{H_ConformalI}).
It is found that the entanglement entropy evolves as $S_A(t)\simeq\frac{c_{\text{eff}}}{3}\log t+\mathrm{const.}$ (see Fig.\ref{4plot}).

\textit{(d) Inhomogeneous quench}

In this case, the initial state is still chosen as the ground state of a massive fermion chain. But now the mass term is position
dependent and has the form [see also Eq.(\ref{Inhomo_quench})]
$
m(x)^{-1}=\sin\beta_0\cdot \sqrt{\Lambda^2+\left(\frac{x-L}{\cos\beta_0}\right)^2},
$
where the parameters $\beta_0\ll 1$ and $\Lambda$ are fixed. Then at time $t=0$, the initial state $|\psi_0\rangle$
evolves according to the hamiltonian in Eq.(\ref{H_ConformalI}). One can find the entanglement entropy
evolves as follows
\be
S_A(t)=\frac{\pi c_{\text{eff}}}{12\beta_0'}\log \left(
\sqrt{1+\frac{t^2}{(\Lambda')^2}}+\frac{t}{\Lambda'}
\right)+\text{const.},
\ee
where the fitting parameters $\beta_0'$ and $\Lambda'$ depend on the mass term $m(x)$, but are independent of
the transmission parameter $\lambda$. 
That is, the entanglement entropy is fitted by $c_{\text{eff}}$, $\beta_0'$ and $\Lambda'$, but only 
$c_{\text{eff}}$ is a $\lambda$ dependent fitting parameter.
In this way, we can obtain $c_{\text{eff}}(\lambda)/c(\lambda=1)$
in Fig.\ref{Compare}.

For all the cases above, by fitting the entanglement entropy in Fig.\ref{4plot} for different $\lambda$,
we obtain $c_{\text{eff}}(\lambda)/c(\lambda=1)$
in Fig.\ref{Compare}. Remarkably, for both equilibrium and non-equilibrium cases, $c_{\text{eff}}(\lambda)/c(\lambda=1)$
fall on the same curve, which agrees with the CFT analysis.

In addition, we compare the numerical results of $c_{\text{eff}}(\lambda)$ with the analytical result
which was obtained in the ground-state case,\cite{Eisler1005}
with the following expression
\be\label{AnalyticalCeff}
c_{\text{eff}}(\lambda)=\frac{12}{\pi^2}\cdot I(\lambda),
\ee
where
$
I(\lambda)=-\frac{1}{2}\Big\{
\big[(1+\lambda)\ln (1+\lambda)+(1-\lambda)\ln (1-\lambda)\big]\ln \lambda+(1+\lambda)\text{Li}_2(-\lambda)+(1-\lambda)\text{Li}_2(\lambda)\Big\},
$
and $\text{Li}_2(\lambda)$ is the dilogarithm function defined by
$
\text{Li}_2(\lambda)=-\int_0^{\lambda}\frac{\ln(1-x)}{x}dx.
$
As shown in Fig.\ref{Compare}, the numerical results agree with the analytical result in an excellent way.


\section{Quasi-particle picture and its application}
\label{Sec_Quasiparticle}

In the previous discussion, we focus on cases when the conformal interface is along the
entanglement cut. Then one can use conformal mappings to reach the configuration
in Fig.\ref{AfterMap}, the partition function
on which can be easily evaluated (after taking periodic boundary condition).
But there are still cases it is not clear how to solve with conformal mappings: (i) When the
conformal interface is inside (or outside) the subsystem $A$, by mapping $\rho_A$ to a strip,
the conformal interface is no longer a straight line. It is difficult to evaluate the partition function
in this case.
(ii) For a more generic quantum quench such that $\beta(x)$ in $|\phi_0\rangle$
is an arbitrary function of $x$, even if the conformal interface is along the entanglement cut, we are not sure
how to find a conformal mapping to get a simple configuration, \textit{e.g.}, Fig.\ref{AfterMap}.
In these cases, it will be helpful to consider the quasi-particle picture,\cite{G_Quench1,G_Quench2,Quench_Review}
which assumes the initial state as the source of EPR pairs.
Starting from $t=0$, these EPR pairs, which carry entanglement, move in opposite directions with light speed $c=1$.
At time $t$, entanglement between two regions with distance $d=t$ starts to be created.
In the following, we will apply this quasiparticle picture to two interesting examples.

\begin{figure}[ht]
\includegraphics[width=3.2in]{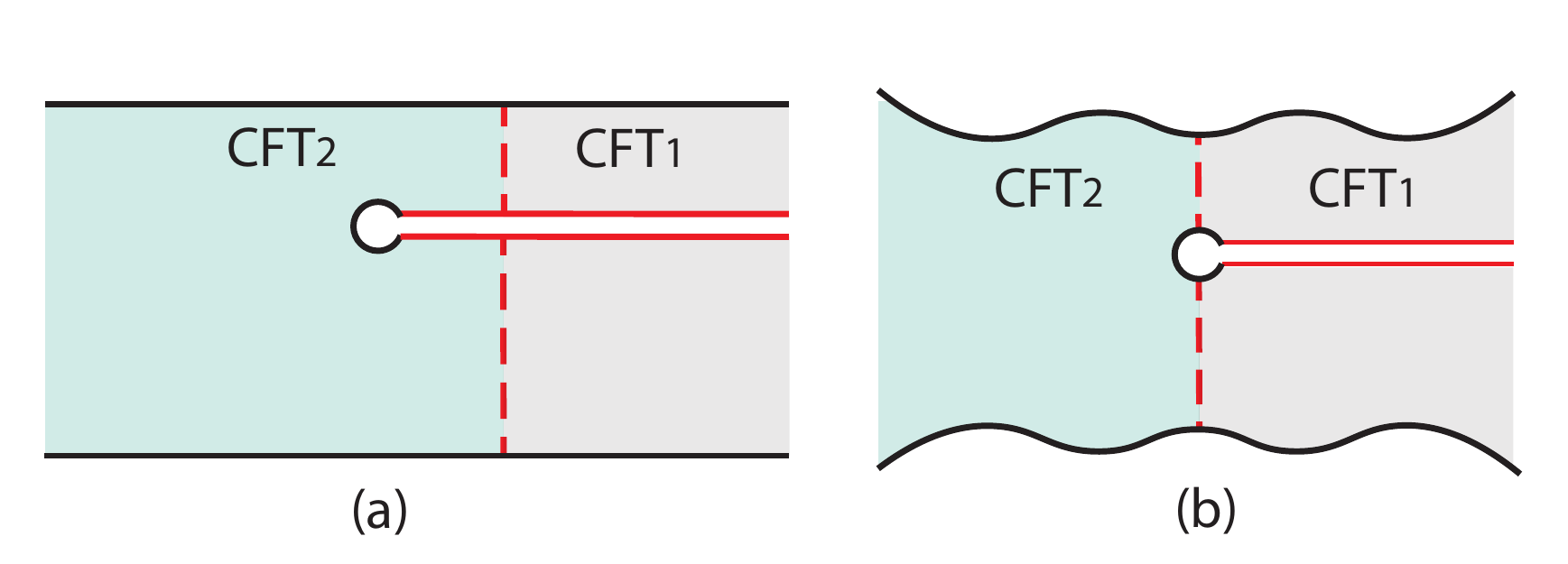}
\caption{
Setup for (a) a global quench with a conformal interface inside the subsystem $A=(0,\infty)$, and (b)
an arbitrary inhomogeneous quantum quench with a conformal interface along the entanglement cut.
}\label{GlobalInsideS}
\end{figure}

\subsection{ A global quench with a conformal interface inside the subsystem}
\label{Global_inside}

As a warm up, let us first consider the case in Fig.\ref{Interface} (b), \textit{i.e.}, a global quench with a
conformal interface along the entanglement cut. Different from the case without a conformal interface,
when quasiparticles hit the conformal interface,
only part of them will transmit. Let us denote the transmission coefficient as $\mathcal{T}$, then the entanglement
entropy $S_A(t)$ for $A=(0,\infty)$ has the form
\be
S_A(t)=\mathcal{T}\int_{-t}^t \rho(x) dx,
\ee
where $\rho(x)$ may be viewed as the density of EPR pairs that carry entanglement.
For the global quench we studied here,
$\rho(x)$ is a constant and may be chosen as \cite{Quench_Review}
\be\label{LR_density0}
\rho(x)=\frac{\pi c}{6\beta},
\ee
Then one has
\be\label{Global001a}
S_A(t)=\frac{\pi c \mathcal{T}}{3\beta}t.
\ee

A remark here: $\rho(x)$ may be alternatively viewed as the left-right entanglement density, which
serves as entanglement sources for the real space entanglement after a quantum quench (See the
discussion in Appendix.\ref{Sec_Appendix_left_right} for more details.). The basic picture is as follows:
Given the regularized conformal boundary state $|\phi_0\rangle$ in Eq.(\ref{InitialState}), there is
entanglement between the left-movers and right-movers. When the system is quenched to a
critical point at $t=0$, the left-movers and right-movers propagate in opposite directions in space,
which results in real-space entanglement. In other words, the real-space entanglement after a quantum quench
originates from the left-right entanglement in the initial state.

By comparing Eq.(\ref{Global001a}) with the result in Eq.(\ref{SAt_Global}), one can find the transmission coefficient as
\be\label{TransmissionC}
\mathcal{T}=\frac{c_{\text{eff}}}{c}.
\ee
Now let us study the case where the conformal interface is inside the subsystem $A$, as shown in Fig.\ref{GlobalInsideS} (a).
It is straightforward to express the entanglement entropy $S_A(t)$ as follows
\be
\begin{split}
S_A(t)=&\Theta(d-t)\int_{-t}^t\rho(x)dx+\Theta(t-d)\cdot \mathcal{T}\int_{2d-t}^t\rho(x)dx\\
&+\Theta(t-d)\int_{-t}^{-t+2d}\rho(x)dx,
\end{split}
\ee
where $\Theta(x)=1$ for $x>0$ and $0$ for $x<0$. Considering $\rho(x)=\frac{\pi c}{6\beta}$ in Eq.(\ref{LR_density0}),
one can immediately obtain
\be\label{SA_Global_Inside}
S_A(t)=\left\{
\begin{split}
&\frac{\pi c}{3\beta}t,  \quad & t<d,\\
&\frac{\pi c}{3\beta}d+\frac{\pi c_{\text{eff}}}{3\beta}(t-d), \quad & t>d.
\end{split}
\right.
\ee
That is, for $t<d$, $S_A(t)$ grows linearly in time with slope $\frac{\pi c}{3\beta}$; for $t>d$,
$S_A(t)$ also grows linearly in time but with a slope $\frac{\pi c_{\text{eff}}}{3\beta}$.
In the limit $c_{\text{eff}}=c$, \textit{i.e.}, the interface is totally transmissive,
one can find $S_A(t)=\frac{\pi c}{3\beta}t$ for arbitrary $t$, as expected.
In the other limit $c_{\text{eff}}=0$, \textit{i.e.}, there is a ``wall" at $x=d$, then one has
$S_A(t>d)=\frac{\pi c}{3\beta}d$, which is saturated, also as expected.\cite{Quench_Review}

\begin{figure}[ht]
\includegraphics[width=2.5in]{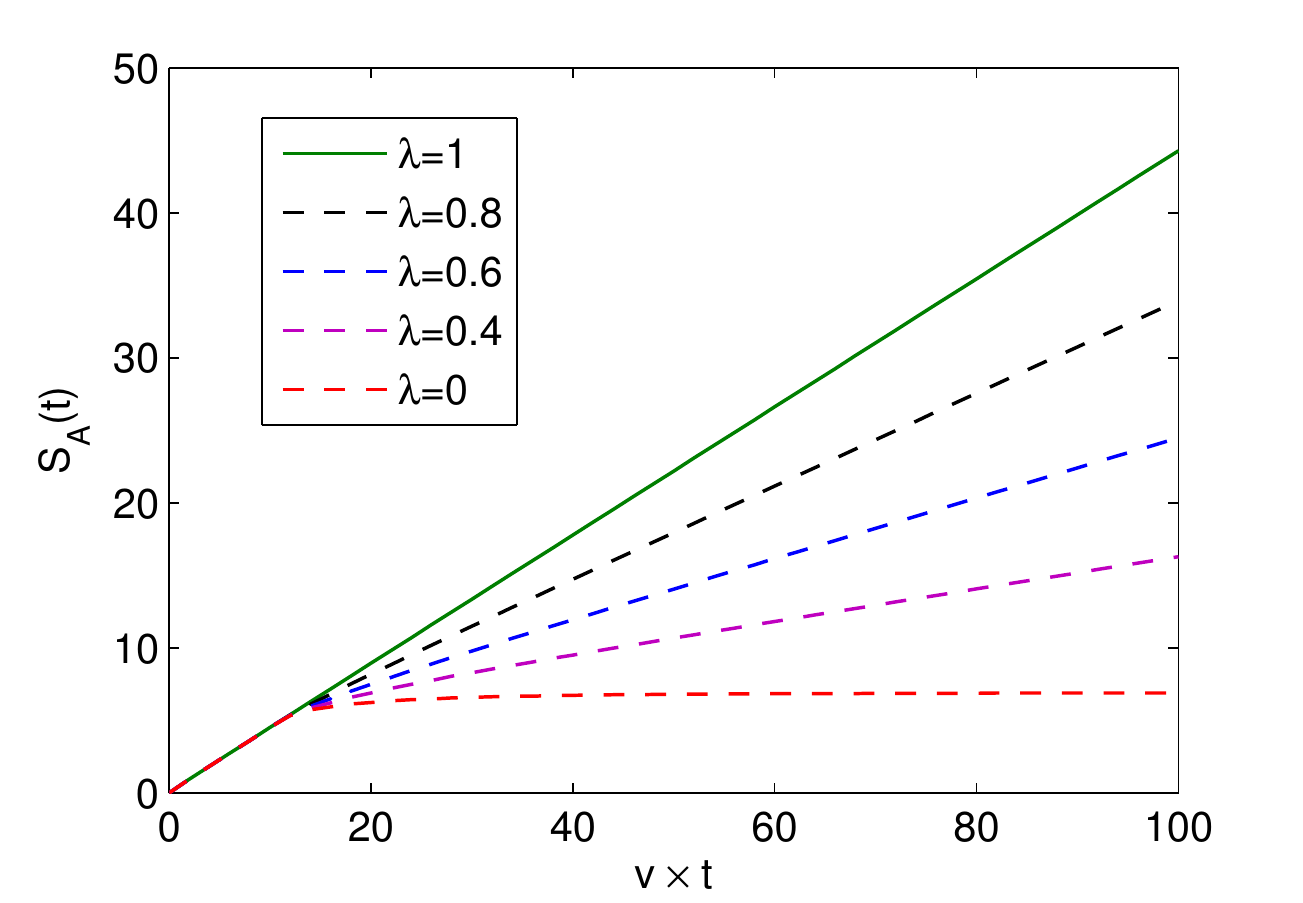}
\caption{
Numerical study of $S_A(t)$ after a global quench. The subsystem $A$ is chosen as $A=[L+1,2L]$, and the conformal interface
is located at $[L+d, L+d+1]$. Here we choose $L=300$ and $d=10$.
}\label{GlobalInside}
\end{figure}

To confirm the behavior of $S_A(t)$ in Eq.(\ref{SA_Global_Inside}) numerically, now
we consider a free fermion chain after a global quantum quench with the conformal interface located
at $[L+d,L+d+1]$, while the subsystem $A$ is still chosen as $A=[L+1,2L]$.
\textit{i.e.}, the conformal interface is no longer located at the entangling cut, but has
a distance $d$ from it. As shown in Fig.\ref{GlobalInside},
it is observed that $S_A(t)$ grows linearly in time for both $vt<d$ and $vt>d$,
where $v$ is the group velocity of quasi-particles in the low energy limit and has the value $v=2$ here.
The difference is that the slope of $S_A(t)$ is $\propto c$ for $vt<d$ but $\propto c_{\text{eff}}$ for $vt>d$,
in agreement with the quasi-particle picture in Eq.(\ref{SA_Global_Inside}).

\subsection{An arbitrary quantum quench based on
quasi-particle picture}
\label{Conjecture}

Now we consider a quantum quench with arbitrary $\beta(x)$ in the initial state
$|\phi_0\rangle$ in Eq.(\ref{phi0_inhom}) (see Fig.\ref{GlobalInsideS} (b) for example),
and the conformal interface is along the entanglement cut.
We will show that, based on the quasi-particle picture, the entanglement
evolution for an arbitrary quench is also suppressed by a factor $c_{\text{eff}}/c$.

First, we make an assumption that the transmission coefficient $\mathcal{T}$
of the conformal interface is independent of the EPR-pair density $\rho(x)$, and has the same
form as $\mathcal{T}$ in a global quench in Eq.(\ref{TransmissionC}).
This assumption is made based on the solvable inhomogeneous quantum quench in Sec.\ref{QuenchCases},
as follows.

Considering the configuration in Fig.\ref{Interface} (d), the entanglement entropy of
subsystem $A=(0,\infty)$ can be expressed as
\be\label{Inhomo_S}
\begin{split}
S_A(t)&=\int_{-t}^t\mathcal{T}(x) \rho(x)dx,
\end{split}
\ee
where the EPR-pair density $\rho(x)$ is not homogeneous and has the form \cite{Sot0808} (see also Appendix \ref{Sec_Appendix_left_right})
\be
\rho(x)=\frac{\pi c}{6\beta(x)},
\ee
where $\beta(x)$ is given in Eq.(\ref{Inhomo_quench}), and is approximated as
$\beta(x)\simeq 4\beta_0\sqrt{\Lambda^2+x^2}$ by considering $\beta_0\ll 1$.
Suppose that the transmission coefficient $\mathcal{T}$ is independent
of EPR-pair density $\rho(x)$, then $S_A(t)$ can be written as
\be
\begin{split}
S_A(t)&=\mathcal{T}\int_{-t}^t\frac{\pi c}{24\beta_0}\frac{1}{\sqrt{\Lambda^2+x^2}}dx\\
&=\mathcal{T}\cdot \frac{\pi c}{12\beta_0}\cdot \log \left(
\sqrt{1+\frac{t^2}{\Lambda^2}}+\frac{t}{\Lambda}\right).
\end{split}
\ee
By comparing with the expression of $S_A(t)$ in Eq.(\ref{SA_inhomogeneous}),
one can find that $\mathcal{T}=c_{\text{eff}}/c$, which is the same as the
global quench case in Eq.(\ref{TransmissionC}).
From this inhomogeneous-quench example, we conjecture that the transmission coefficient
$\mathcal{T}$ is independent of EPR-pair density,
and has the universal value in Eq.(\ref{TransmissionC}).

Then we can move on to an arbitrary quantum quench with a conformal interface
along the entanglement cut. Here `arbitrary' means $\beta(x)$ in the initial state $|\phi_0\rangle$ in Eq.(\ref{phi0_inhom})
is an arbitrary function which smoothly varies with $x$, and so is the EPR-pair density $\rho(x)$.
Then based on quasi-particle picture, the entanglement entropy $S_A(t)$ may be written as
\be\label{ArbitraryS}
\begin{split}
S_A(t)&=\mathcal{T}\int_{-t}^t\rho(x)dx
=\frac{c_{\text{eff}}}{c}S_{A,0}(t),
\end{split}
\ee
where $S_{A,0}(t)$ denotes the entanglement entropy without a conformal interface.

We emphasize that Eq.(\ref{ArbitraryS}) is obtained based on the quasi-particle picture
and the assumption that $\mathcal{T}=c_{\text{eff}}/c$ for arbitrary quantum quenches.
Therefore, the result in Eq.(\ref{ArbitraryS}) is a conjecture, although
we have numerically checked cases with different $\beta(x)$
in a free-fermion chain to confirm it.\cite{Wen_unpublished}
A more strict proof with conformal mapping (or other methods) is still desirable.
In addition, it will be interesting to connect our results with quantum quenches
in higher dimensions, where the ``entanglement tsunami" carries entanglement from the boundary.\cite{Liu2014,CasiniLiu}

\section{Concluding remarks}
\label{Sec_conclusion}

In this work, by using conformal mappings, we show that the leading term of
entanglement entropy after a quantum quench is suppressed by a conformal
interface in the same way as that in the ground-state case.
We study three different quantum quenches explicitly, including a global quench,
a local quench, and a homogeneous quench.
For each case, the effect of conformal interface in the entanglement entropy evolution
is to replace the central charge $c$ with $c_{\text{eff}}$. Here $c_{\text{eff}}$
is the same as that in the ground-state case.
Our conclusion is confirmed by numerical calculations based on a free fermion chain.
In addition, based on the quasi-particle picture, we conjecture that our conclusion
holds for an arbitrary
quantum quench whose initial state can be described by a regularized conformal
boundary state.

Although our discussion mainly focuses on two CFTs connected by a conformal interface,
it is straightforward to generalized our conclusion to several CFTs joining at a conformal
interface/junction.\cite{Miller1701}

There are many interesting future problems to study, and we mention some of them here.

-- The conformal interface studied in this work is located along the entanglement cut (except for the example
on quasi-particle picture in Sec.\ref{Global_inside}).
As far as we know, even for the equilibrium case, the entanglement entropy with a
conformal interface inside (or outside) the subsystem is not studied yet.
It is our future work to understand how a conformal interface inside (or outside)
the subsystem affects the entanglement entropy
for both equilibrium and non-equilibrium cases.

-- Another way to study entanglement entropy in CFTs is based on the correlation function of
twist operators.\cite{CC_2004, Quench_Review} It is interesting to study correlation functions
of field operators in the presence of conformal interfaces.
This may provide a good way to study the case that the conformal interface is inside (or outside)
the subsystem.

-- Recently, the Loschmidt echo and bipartite fidelity of free-boson CFTs after a local quench (by gluing two CFTs with
a conformal interface) was studied.\cite{Zhou1706} It is interesting to study the relation between the dynamics therein
with the time evolution of entanglement entropy in our work.

-- It is also interesting to study the holographic entanglement evolution
after quantum quenches in the presence of a conformal interface.

\section{Acknowledgement}

We thank Andreas W. W. Ludwig ,
Tokiro Numasawa, and Tomonori Ugajin for helpful discussions and collaborations on related works.
In particular, XW thanks John Cardy for his helpful interpretation on the work \onlinecite{Cardy1608}.
We are grateful to the KITP Program “Quantum Physics of Information”  (Sep 18 - Dec 15, 2017). 
XW is supported by the postdoc fellowship from
Gordon and Betty Moore Foundation EPiQS Initiative through Grant No.~GBMF4303 at MIT.
YW is supported by the Gordon and Betty Moore Foundation’s EPiQS Initiative through
Grant No. GBMF4305 at the University of Illinois.
This work was supported by the NSF under Grants No. NSF PHY-1125915 (SR).

\appendix

\section{Effect of boundary condition on entanglement entropy}
\label{Effect_BC}

\begin{figure}[htp]
\includegraphics[width=2.50in]{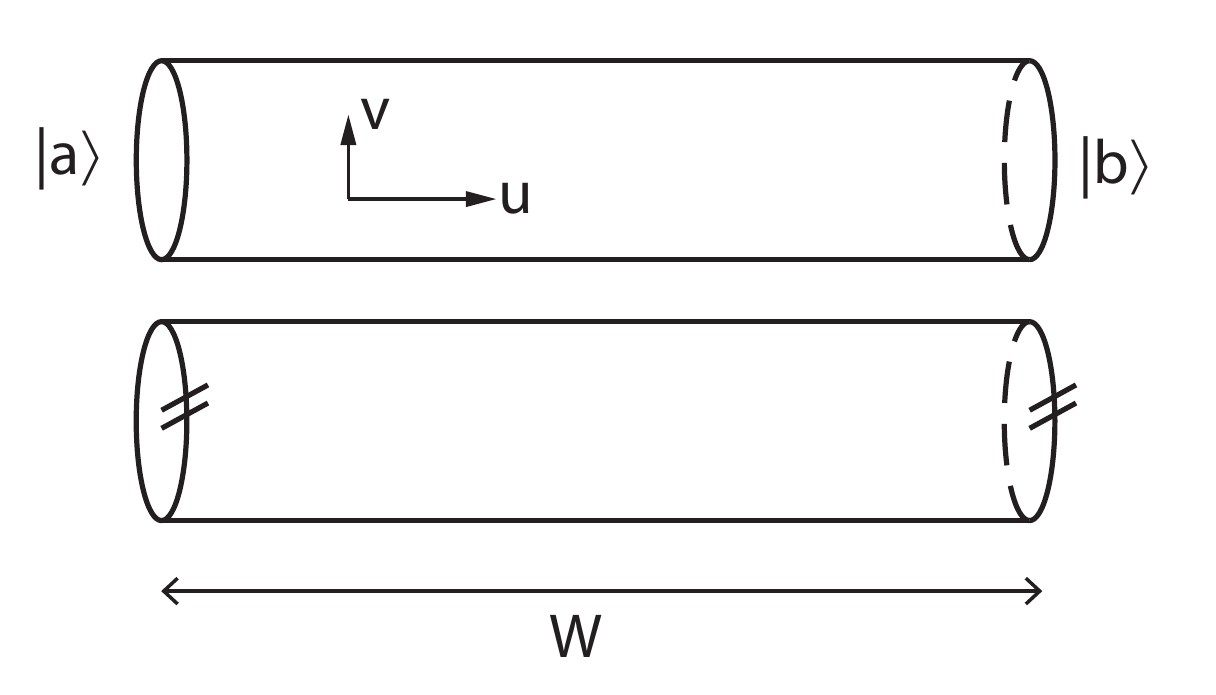}
\caption{Configuration for $\text{tr}\left(\rho_A^n\right)$ without conformal interfaces.
The circumference in $v=\text{Im}\,w$ direction is $2\pi n$,
and the width in $u=\text{Re}\,w$ direction is $W$.
For the cylinder geometry, we have two conformal
boundary conditions $|a\rangle$ and $|b\rangle$ on the two edges.
And the torus is a cylinder with periodic boundary condition.
}\label{cylinder_torus}
\end{figure}

In this appendix, we argue that the boundary conditions $|a_{1(2)}\rangle$ and
$|b_{1(2)}\rangle$ in Fig.\ref{AfterMap} does not affect the leading term in entanglement entropy.
For simplicity, let us first consider a homogeneous CFT, and then include
conformal interfaces later. The discussion follows closely with that in Ref.\onlinecite{Cardy1608}.

Let us start with the strip (without conformal interfaces), which represents the
reduced density matrix $\rho_A$, in Fig.\ref{AfterMap}.
By taking $n$ copies of strips
and considering the trace operation, \textit{i.e.},
$\text{Tr}\left(\rho_A^n\right)$, one obtains a cylinder
with two open boundary conditions $|a\rangle$ and $|b\rangle$ (see Fig.\ref{cylinder_torus}).
To make a comparison, we also consider a cylinder with periodic boundary condition, \textit{i.e.}, a torus
(see Fig.\ref{cylinder_torus}).
The circumference in $v=\text{Im}\,w$ direction of the cylinder/torus is $2n\pi$,
and the length in $u=\text{Re}\,w$ direction is denoted as $W$.
Since $W\gg 1$, it is convenient to consider the partition function as the path integral
for a CFT on a circle of circumference $2n\pi$, propagating along $u$-direction with imaginary time $W$.
Then one has\cite{Cardy1608}
\be
\left\{
\begin{split}
\text{Tr}\left(\rho_A\right)&=Z_1=
\tilde{q}^{-c/24}\sum_{k}\langle a|k\rangle\langle k|b\rangle \tilde{q}^{\delta_k}\\
\text{Tr}\left(\rho_A^n\right)&=Z_n=
\tilde{q}^{-c/24n}\sum_{k}\langle a|k\rangle\langle k|b\rangle \tilde{q}^{\delta_k/n}\\
\end{split}
\right.
\ee
for the cylinder, and
\be
\left\{
\begin{split}
\text{Tr}\left(\rho_A\right)&=Z_1=
\tilde{q}^{-c/24}\sum_{k}d_k \tilde{q}^{\delta_k}\\
\text{Tr}\left(\rho_A^n\right)&=Z_n=
\tilde{q}^{-c/24n}\sum_{k}d_k \tilde{q}^{\delta_k/n}\\
\end{split}
\right.
\ee
for the torus. Here $\tilde{q}=e^{-2W}$, $\delta_k$ are dimensions of bulk operators, and
the positive integers $d_k$ are degeneracy factors.
Then $\text{tr}\left(\rho_A^n\right)$ after normalization can be written as
$
\text{tr}\,\rho_A^n=\frac{Z_n}{Z_1^n}.
$
In the limit $W\gg 1$, $\text{tr}\left(\rho_A^n\right)$ is approximated as
\be
\text{tr}\left(\rho_A^n\right)\simeq\frac{
\langle a|0\rangle\langle 0|b\rangle \tilde{q}^{-c/24n}
}{
(\langle a|0\rangle\langle 0|b\rangle )^n\tilde{q}^{-cn/24}
},
\ee
for a cylinder, and
\be
\text{tr}\left(\rho_A^n\right)\simeq\frac{
 \tilde{q}^{-c/24n}
}{
\tilde{q}^{-cn/24}
},
\ee
for a torus.
Then the $n$-th Renyi entropy can be expressed as
\be
S_A^{(n)}\simeq \frac{c}{12}\left(1+\frac{1}{n}\right)W-g_a-g_b,
\ee
for a cylinder, where $g_{a,b}=-\log\langle a,b|0\rangle$ are
the Affleck-Ludwig boundary entropies.\cite{Affleck_Ludwig}
For the torus, one has
\be
S_A^{(n)}\simeq \frac{c}{12}\left(1+\frac{1}{n}\right)W.
\ee
Considering $W\gg 1$, the leading term of Renyi entropy proportional to $W$
are the same for a cylinder and a torus.
The difference only happens for the subleading term.

In the above analysis, what is essential is that in the limit $W\gg 1$, only the ground state $|0\rangle$
dominates in the partition function, and the boundary states contribute to a finite
overlap $\langle a|0\rangle$ ($\langle b|0\rangle$).
Now we include conformal interfaces in $\text{Tr}(\rho_A^n)$.
Again, we consider the partition function as the path integral for a CFT living on a circle of length $2n\pi$,
propagating along $\text{Re}\,w$ direction. Denoting the hamiltonian on the circle as $H_{\text{CFT}}$,
then one has
$
\text{Tr}(\rho_A^n)=\sum_k\langle A|k\rangle\langle k|B\rangle \langle k|e^{-WH_{\text{CFT}}}|k\rangle
$
for the cylinder geometry, and
$
\text{Tr}(\rho_A^n)=\sum_k \langle k|e^{-WH_{\text{CFT}}}|k\rangle
$
for the torus geometry, where $|k\rangle$ span the complete bases in the Hilbert space.
Here $|A\rangle$ ($|B\rangle$) denotes the boundary condition on the left (right) edge of the cylinder, with
$|A\rangle=|a_1\rangle$ for $v\in [-\frac{\pi}{2}+2n\pi, \frac{\pi}{2}+2n\pi]$ and
$|A\rangle=|a_2\rangle$ for $v\in [\frac{\pi}{2}+2n\pi, \frac{3\pi}{2}+2n\pi]$, and similarly for $|B\rangle$.
In the limit $W\gg 1$, the ground state $|0\rangle$ dominates in the partition function. Then one has
$
\text{Tr}(\rho_A^n)\simeq\langle A|0\rangle\langle 0|B\rangle \langle 0|e^{-WH_{\text{CFT}}}|0\rangle
$
for a cylinder, and
$
\text{Tr}(\rho_A^n)\simeq\langle 0|e^{-WH_{\text{CFT}}}|0\rangle
$ for a torus.
Similar to the case without conformal interfaces, the boundary conditions $|A\rangle$ and $|B\rangle$
only contribute to a finite constant to the entanglement entropy. The leading term of entanglement entropy
is contributed by $\langle 0|e^{-WH_{\text{CFT}}}|0\rangle$, which is the same for both the cylinder
geometry and torus geometry.
Therefore, we conclude that the boundary conditions
$|a_{1(2)}\rangle$ and $|b_{1(2)}\rangle$ in Fig.\ref{AfterMap} have no
contribution to the leading term in entanglement entropy.

\section{Left-right entanglement as sources of real-space entanglement after a quantum quench}
\label{Sec_Appendix_left_right}

Left-right entanglement of a conformal boundary state has been studied recently,\cite{LR_Zayas1407,LR_Das1504}
and applied to the real-space entanglement entropy in Chern-Simons theories.\cite{TEE,Qi1103,LR_Wen,Fliss1705}
Here we propose that the left-right entanglement density of a conformal boundary state
may be considered as the entanglement sources for real-space entanglement in a CFT after a quantum quench.
An intuitive picture is that the entanglement evolution after a quantum quench is introduced by the propagation
of left-moving and right-moving quasiparticles. Tracing back to $t=0$, the entanglement between left-moving
and right-moving quasiparticles must come from the entanglement between left-moving and right-moving modes
of the regularized conformal boundary condition.

Let us define the left-right entanglement density in the following. Now we consider
a regularized conformal boundary state
\be
|\mathfrak{b}\rangle=e^{-\beta H_{\text{CFT}}/4}|b\rangle,
\ee
where $|b\rangle$ is the conformal boundary state defined along a circle of length $L$.
It is noted that $|b\rangle$ is a superposition of Ishibashi states $|h_a\rangle\rangle$,\cite{CardyBC}
which may be explicitly written as
\be
|h_a\rangle\rangle\equiv \sum_{N=0}^{\infty}\sum_{j=1}^{d_{h_a}(N)}|h_a,N;j\rangle
\otimes \overline{|h_a,N;j\rangle},
\ee
where $|h_a,N;j\rangle$ represent the left movers, and $\overline{|h_a,N;j\rangle}$ represent
the right movers. Here $a$ denotes the primary field, and $d_{h_a}(N)$
denotes the dimension of subspace for level $N$ of the conformal family.
Without loss of generality, one can trace over the right-movers, and obtain the reduced density matrix for the left-movers:
\be
\rho_L=\text{tr}_R\left(|\mathfrak{b}\rangle\langle \mathfrak{b}|\right).
\ee
Then one can obtain the entanglement entropy for the left-movers (or right-movers) as follows:\cite{LR_Zayas1407,LR_Das1504}
\be
S=\frac{\pi c}{6\beta}\cdot L+\text{const.}.
\ee
Based on the leading term in $S$ we can define the left-right entanglement density $\rho=S/L$ with the form
\be\label{rhoX1_appendix}
\rho(x)=\frac{\pi c}{6\beta},
\ee
which is nothing but Eq.(\ref{LR_density0}). For inhomogeneous quantum quench with smoothly varying
initial boundary conditions, $\beta(x)$ is position dependent [see Eq.(\ref{phi0_inhom})],
and $\rho(x)$ may be expressed as\cite{left_right_comment}
\be\label{rhoX2}
\rho(x)=\frac{\pi c}{6\beta(x)}.
\ee
We will check several nontrivial examples based on this left-right entanglement density.

One simple example is the global quantum quench with the setup shown in Fig.\ref{Interface} (b), but with no
conformal interface. For subsystem $A=(0,\infty)$, the entanglement entropy has the form
$
S_A(t)=\int_{-t}^t\rho(x)dx=\frac{\pi c}{3\beta}\cdot t,
$
where we have used the definition in Eq.(\ref{rhoX1_appendix}).

Another interesting example is the inhomogeneous quantum quench discussed in the main text, with $\beta(x)$
shown in Eq.(\ref{Inhomo_quench}). For simplicity, we do not include the conformal interface here.
Considering that $\beta_0\ll 1$, $\beta(x)$ in Eq.(\ref{Inhomo_quench}) can be approximated as
$
\beta(x)\simeq 4\beta_0\sqrt{\Lambda^2+x^2}.
$
Then the entanglement entropy for $A=(0,\infty)$ has the form
$
S_A(t)\simeq \int_{-t}^t\rho(x)dx
=\frac{\pi c}{12\beta_0}\int_0^{t}\frac{1}{\sqrt{\Lambda^2+x^2}}dx
=
\frac{\pi c}{12\beta_0} \log\left(\frac{t}{\Lambda}+\sqrt{1+\frac{t^2}{\Lambda^2}}\right),
$
which agrees with the result in Ref.\onlinecite{Inhomo_Quench}.
Furthermore, it is also interesting to check the case with $A=[l,\infty)$ where $l\gg \Lambda$.
In Ref.\onlinecite{Inhomo_Quench}, it has been found that
\be\label{S_At001}
S_A(t)\simeq
\left\{
\begin{split}
&\frac{c}{6}\log l+\frac{\pi c}{12\beta_0}\cdot \frac{t}{l}, \quad t\ll l,\\
&\frac{\pi c}{24\beta_0}\log (t^2-l^2), \quad t>l.
\end{split}
\right.
\ee
Note that for $t=0$, one has $S_A(t=0)=\frac{c}{6}\log l$, which is contributed by the initial state.
Here we are interested in the excess time-dependent part contributed by quench.
Then one can express the entanglement entropy as
$
S_A(t)=\int_{l-t}^{l+t}s(x)dx.
$
For $t\ll l$, after some simple algebra, one can find
$
S_A(t)\simeq \frac{\pi c}{24\beta_0}\log \frac{l+t}{l-t}\simeq \frac{\pi c}{12\beta_0}\cdot \frac{t}{l},
$
which is nothing but the time dependent term in Eq.(\ref{S_At001}).
For $t>l$, one has
$
S_A(t)=\frac{\pi c}{24\epsilon_0}\log\left(
\frac{t+l}{\Lambda}+\sqrt{1+\frac{(t+l)^2}{\Lambda^2}}
\right)-
\frac{\pi c}{24\epsilon_0}\log\left(
-\frac{t-l}{\Lambda}+\sqrt{1+\frac{(t-l)^2}{\Lambda^2}}
\right).
$
Considering $t, l\gg \Lambda$, and ignoring constant terms, $S_A(t)$ can be further simplified as
$
S_A(t)\simeq \frac{\pi c}{24\beta_0}\log (t^2-l^2),
$
which is nothing but the second equation in Eq.(\ref{S_At001}).

The merit of this method is that, for a generic initial state $|\phi_0\rangle$,
 we do not need to find out the conformal mapping to map the reduced density matrix to a simple
configuration, which may be difficult for an arbitrary $\beta(x)$.
Now, what one needs to do is simply calculating an integral $S_A(t)\simeq \int \rho(x)dx$,
where $\rho(x)$ is expressed in Eq.(\ref{rhoX2}).
Numerically, we have checked various choices of $\beta(x)$ in $|\phi_0\rangle$
based on a free fermion chain,
and the results agree very well with the quasi-particle picture with EPR-pair density of
the form in Eq.(\ref{rhoX2}).\cite{Wen_unpublished}

In short, in this Appendix, we propose that the left-right entanglement can be viewed
as the source for entanglement evolution in real space after a quantum quench, as long as
the initial state is described by a regularized conformal boundary state.
This is straightforwardly checked in a global quench. For inhomogeneous quenches,
we conjecture that the left-right entanglement density has the form in Eq.\eqref{rhoX2},
which is verified (but not proved) based on a solvable inhomogeneous quantum quench.

\section{Mass term in the global quench of a free fermion chain}
\label{Sec_GlobalQuench_MassTerm}

In this section, following Refs.\onlinecite{Qi1103} and \onlinecite{Miyaji1412},
we give a brief review on how to fix the parameter $\beta$ in 
the global quantum quench [see Eq.\eqref{InitialState}] from the mass term in a free fermion chain.
In the low energy limit, the massless free fermion is discribed by the Hamitonian
\be
H_{\text{CFT}}=H_R+H_L,
\ee
where 
\be
H_R=\sum_k vkc_k^{\dag}c_k, \quad H_L=-\sum_k vk d_k^{\dag}d_k.
\ee
are the Hamiltonians describing the right- and left-moving massless fermions.
Here, $k$ is momentum, and $c_k$ ($d_k$) are fermionic operators.
Then one can introduce a massive term to gap out the fermions:
\be
H_{RL}=2m\sum_{k} c_k^{\dag}d_k+h.c.
\ee
(The reason why we add a factor `2' can be clearly seen below.)
Then one can find that in the low energy limit $k\to 0$,
the ground state for this massive fermion $H=H_{\text{CFT}}+H_{RL}$ can be written as\cite{{Qi1103}}
\be
|G\rangle\simeq e^{-\frac{1}{4m}H_{\text{CFT}}}|b\rangle,
\ee
where $|b\rangle$ is the conformal boundary state for a free fermion CFT. 
In the global quantum quench as studied in Sec.\ref{Sec_Lattice} (b), we choose the initial state as $|G\rangle$.
Comparing with Eq.\eqref{InitialState}, one can find that 
\be
\beta=\frac{1}{m}.
\ee
That is, the `temperature' in the intitial state is proportional to the mass of the fermion.
(It is noted that in the numerical calculation on a lattice model, $\beta$ is not exactly $1/m$ but needs fine tuning.)


\begin{thebibliography}{99}
\bibitem{TEE}
To be more precise, it is the left-right entanglement of an Ishibashi state (rather than a conformal boundary state
which is a superposition of Ishibashi states) that contributes to the real-space entanglement in Chern-Simons field theories.
\bibitem{left_right_comment}
It should be more convincing if we can find that the left-right entanglement for $|\mathfrak{b}\rangle$
has the form $S\simeq \int_0^L\frac{\pi c}{6\beta(x)}dx$, which we do not know how to calculate at
the current stage.
\end{thebibliography}

\end{document}